\documentclass[conference]{IEEEtran} % [journal]
\IEEEoverridecommandlockouts
\usepackage[utf8]{inputenc}
\usepackage{graphicx}
\usepackage{tabularray}
\usepackage{array}
\usepackage{epsfig}
\usepackage{epstopdf}
\usepackage{lipsum,multicol}
\usepackage{cite}
\usepackage[ruled,vlined,linesnumbered,noresetcount]{algorithm2e}
\usepackage{mathtools}
\usepackage{cuted}
\usepackage{blindtext}
\usepackage{wrapfig}
\usepackage{url}

\usepackage{soul}
\usepackage{xcolor}

\usepackage{longtable}
\usepackage{booktabs}

\usepackage{balance}

\usepackage{enumitem}
\usepackage{amsmath,amssymb}
\usepackage{multirow}

\usepackage[noend]{algpseudocode}
\makeatletter
\def\BState{\State\hskip-\ALG@thistlm}
\makeatother

\usepackage{textcomp}

\usepackage{microtype}

\begin{document}

\title{  Distributed MIMO for 6G sub-Networks in the Unlicensed Spectrum\\

\thanks{The work of M. Elwekeil and S. Buzzi was supported by H2020 Marie Skłodowska-Curie Actions (MSCA) Individual Fellowships (IF) IUCCF, grant 844253. The work of P. Baracca was supported by the German Federal Ministry of   Education and Research (BMBF) project 6G-ANNA grant 16KIS077K.}
}
%\author{\IEEEauthorblock{Mohamed~Elwekeil$^{1,2}$,
%         Lorenzo~Galati~Giordano$^{1}$, Paolo~Baracca$^{1}$, and~Stefano~Buzzi$^{2,3,4}$ } \\ %\vspace{-0.8cm}% <-this % stops a space % Senior Member,~IEEE,
% \IEEEauthorblockA{$^{1}$\textit{Nokia, Germany.} \\ %(email:\{\}@).}% <-this % stops a space
%$^{2}$\textit{University of Cassino and Southern Lazio, Italy.} \\
%$^{3}$\textit{Consorzio Nazionale Interuniversitario per le Telecomunicazioni, Parma, Italy.} \\
%$^{4}$\textit{Politecnico di Milano, Italy.} 
% }
%}
\author{Mohamed~Elwekeil$^{1,2}$,
         Lorenzo~Galati~Giordano$^{1}$, Paolo~Baracca$^{1}$, and~Stefano~Buzzi$^{2,3,4}$ \\ %\vspace{-0.8cm}% <-this % stops a space % Senior Member,~IEEE,
 $^{1}$\textit{Nokia, Germany.} \\ %(email:\{\}@).}% <-this % stops a space
$^{2}$\textit{University of Cassino and Southern Lazio, Italy.} \\
$^{3}$\textit{Consorzio Nazionale Interuniversitario per le Telecomunicazioni, Parma, Italy.} \\
$^{4}$\textit{Politecnico di Milano, Italy.} \\
Email: mohamed.elwekeil@nokia.com, lorenzo.galati\_giordano@nokia-bell-labs.com, \\ paolo.baracca@nokia.com, buzzi@unicas.it
}
%\author{mohamed.elwekeil2010 }
%\date{August 2022}

\maketitle

\begin{abstract}
In this paper, we consider the sixth generation (6G) sub-networks, where hyper reliable low latency communications (HRLLC) requirements are expected to be met.
We focus on a scenario where multiple sub-networks are active in the service area and assess the feasibility of using the 6 GHz unlicensed spectrum to operate such deployment, evaluating the impact of listen before talk (LBT).
Then, we explore the benefits of using distributed multiple input multiple output (MIMO), where the available antennas in every sub-network are distributed over a number of access points (APs). Specifically, we compare different configurations of distributed MIMO with respect to centralized MIMO, where a  single AP with all antennas is located at the center of every sub-network.
%Then, we explore the benefits of using distributed multiple input multiple output (MIMO) by comparing different antenna configurations per access point (AP) against a centralized MIMO with a single AP with all antennas located at the center of each sub-network.
Our simulation results show the effectiveness of employing distributed MIMO for future sub-networks in the 6 GHz unlicensed spectrum when adaptive transmission power reduction (APR) is used.
 \end{abstract}


 \begin{IEEEkeywords}
 6G, in-X sub-networks, hyper reliable low latency communications (HRLLC), distributed MIMO, unlicensed spectrum.
 \end{IEEEkeywords} %\vspace{-0.5cm}

\section{Introduction}
The forthcoming wireless networks are evolving to ensure data connectivity in many fields.
In particular, research for the sixth generation (6G) of mobile networks is targeting new use cases where extreme requirements in terms of latency, reliability and/or data rate are expected to be met in short-range communications \cite{berardinelli2021extreme}.
In such context, the in-X sub-networks concept has been introduced, where the `X' refers to the entity where the sub-network is deployed, e.g., a robot, a car, a vehicle, etc \cite{adeogun2020towards}. Moreover, the International Telecommunication Union (ITU) has recently introduced hyper reliable low latency communications (HRLLC) as one of the six usage scenarios for 6G \cite{itu2030}: HRLLC targets down to 0.1 ms latency and beyond 5 nines reliability, extending fifth generation (5G) ultra reliable low latency communications (URLLC) requirements.

Currently, only a few works in the literature have investigated the topic of 6G in-X sub-networks (e.g., \cite{ adeogun2020towards, adeogun2023wcnc, dong2023wcnc}), and all with a focus on the licensed spectrum. On the other hand, there are many benefits for unlicensed frequency bands, such as easy deployment, affordable devices, inexpensive operation, and license exemption.
The  Federal Communications Commission (FCC) endorsed an unlicensed use of 1200 MHz around the 6 GHz band in America. Meanwhile, the European Commission allowed the unlicensed utilization of 500 MHz around the 6 GHz band \cite{sathya2020standardization}. This unlicensed spectrum band can be exploited by sub-networks \cite{berardinelli2021extreme}. Note also that, although the concept of sub-networks has been introduced in the context of 6G research, it is still not decided which technology will eventually be used for intra-sub-network communications. 

In fact, although it comes natural to consider cellular-based technologies like 5G/6G as the major candidates for supporting reliability and latency constraints, Wi-Fi protocols are progressively introducing features and techniques aiming at significantly improving their performance \cite{galatiWiFi8}. Accordingly, our study on sub-networks in the unlicensed spectrum in this paper is kept at a standard-agnostic level, focusing on the main common challenge associated with the regulations in place when using the unlicensed spectrum, i.e., ensuring strict quality of service (QoS) requirements while conforming with the carrier sense multiple access with collision avoidance (CSMA/CA) mechanism. CSMA/CA is based on the listen before talk (LBT) procedure, wherein every entity senses the transmission channel to ensure that no concurrent transmissions are ongoing. 

Furthermore, the previous works related to sub-networks have considered a centralized access point (AP) deployment, where all the antennas are located at a single AP for each sub-network.
With distributed multiple input multiple output (MIMO), the antennas are deployed as simple APs, that are spread all over the service area \cite{bjornson2019making}. Here, simple AP means an AP that has a limited number of antennas. 
The antenna distribution brings the APs closer to the end users and improves the propagation diversity. So, distributed MIMO can outperform the traditional centralized MIMO in terms of individual users' data rates, especially for cell-edge users. 
Distributed MIMO has been considered in the literature only for large service areas \cite{bjornson2019making, ngo2017cell}, and its effectiveness for sub-networks has not been studied in the literature yet.

This paper investigates the capability of distributed MIMO to provide strict QoS for sub-networks communication when using the unlicensed spectrum around 6 GHz. In this work, we present a framework for operating multiple coexisting sub-networks, where the channel sensing is made at the AP level. We assess the impact of LBT on the performance. In order to overcome LBT limitations for HRLLC traffic, we present an adaptive transmission power reduction (APR) approach to guarantee that all sub-networks can work without the need to defer any transmission. We then benchmark the performance of different levels of distributed MIMO against that of the centralized MIMO. To the best of the authors' knowledge, this is the first work that considers the utilization of both distributed MIMO and unlicensed spectrum to provide extreme QoS for the 6G sub-networks.

\begin{figure*}[!t]
\begin{center}
\includegraphics[width =1.5\columnwidth]{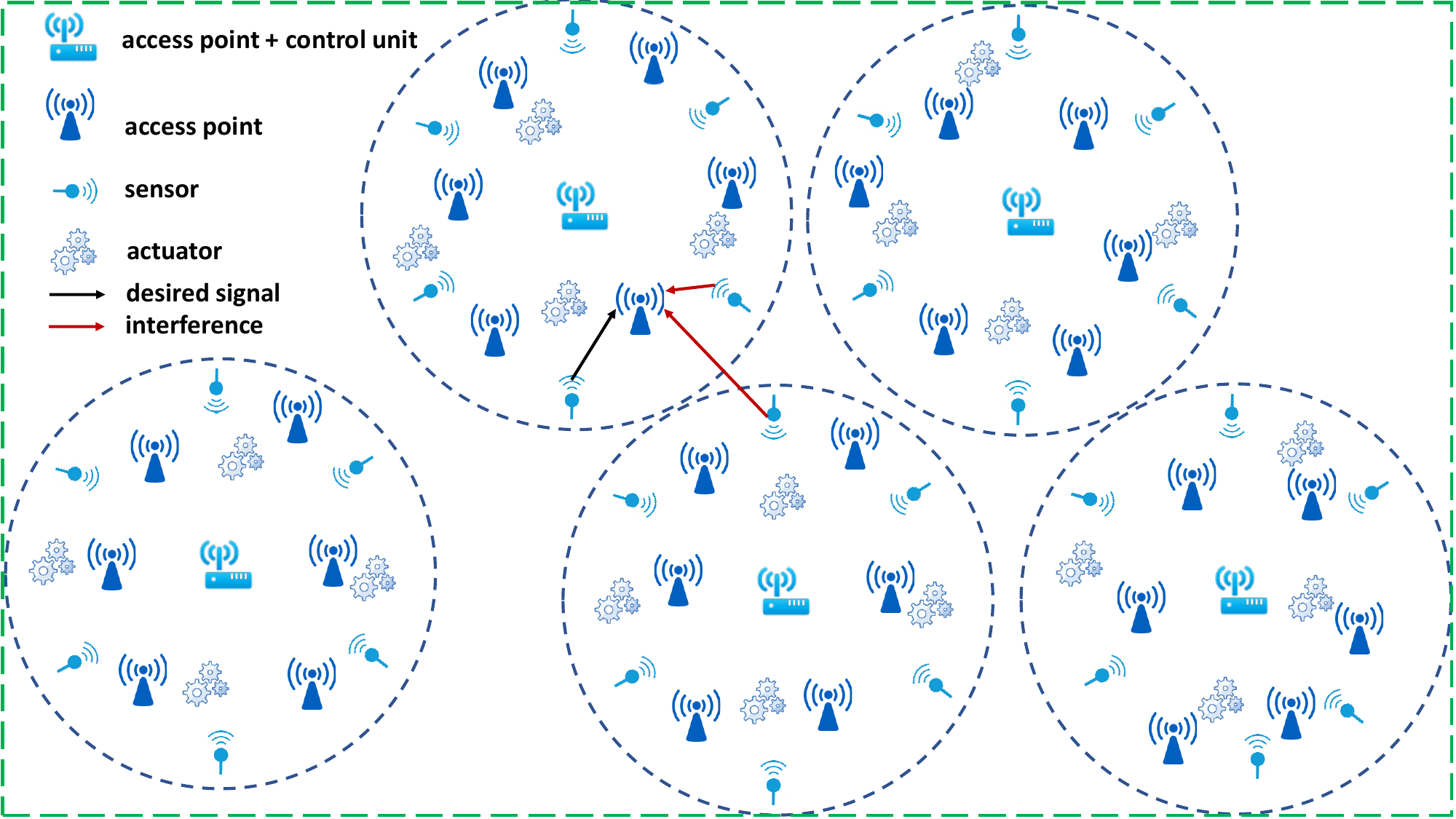}
\caption{ The considered system model of multiple sub-networks, each employing distributed MIMO.}%\vspace{-.2in}
\label{fig:system model}
\end{center}
\end{figure*}
\section{System Model }
The considered system model is shown in Fig. \ref{fig:system model}, where there are $B$ sub-networks operating over the same bandwidth $W$ in the unlicensed spectrum at 6 GHz.
We use $\mathbb{B}=\{BN_1,BN_2,..,BN_b,..,BN_{B}\}$ to denote the set of all the available sub-networks, where the subscript $b$ is the sub-network index, and $B$ is the cardinality of set $\mathbb{B}$.
Thus, the set of the sub-networks indices can be represented by $\mathbb{I}=\{1,2,..,b,..,B\}$.
Every sub-network includes sensors, actuators, a number of distributed APs and a central AP equipped with a control unit (CU). In this paper, we concentrate on the uplink transmissions i.e., from the sensors to the CU through the APs. In such an uplink scenario,  the CU collects the channel state information of the sensors through the distributed APs, and the LBT results made at each AP.
The set of all CUs is represented by $\mathbb{C}=\{CU_1,..,CU_b,..,CU_{B}\}$, where $CU_b$ is the control unit in the sub-network $BN_b$, and the number of employed CUs equals the number of existing sub-networks, $B$. Thus, the set $\mathbb{I}$ also represents the set of the indices of all CUs in the system.

We assume that each sensor is equipped with a single antenna to transmit the packets to the corresponding CU through the available APs in its sub-network.
Here, we consider an ideal wired connection between the CU and the corresponding APs.
Furthermore, we assume a total number of $M$ antennas employed by the set of APs in each sub-network.
These $M$ antennas are equally distributed among the $A_b$ APs available in the sub-network $BN_b$.
Thus, the number of antennas at every $AP$ is $M_{a_b} = \frac{M}{A_b}$. 

The sub-network $BN_b$ contains a set of sensors whose indices are collected in the set $\mathbb{O}_b=\{o_{1,b},..,o_{o_b,b},..,o_{O_b,b}\}$, and a set of APs whose indices are collected in the set $\mathbb{A}_b=\{a_{1,b},..,a_{a_b,b},..,a_{A_b,b}\}$. Here,  $O_b$ and $A_b$ are the numbers of sensors and APs in the sub-network $b$, respectively. Furthermore, the subscripts $o_b$, and $a_b$ are the corresponding indices of sensors and APs within the sub-network $b$, respectively. Moreover, $o_{o_b,b}$, and $a_{a_b,b}$ are the corresponding unique indices of a certain sensor and a certain AP, respectively, within the whole considered system including all sub-networks. Specifically, we define the indices within the overall system such that $o_{o_b,b} = o_{O_{b-1},b-1} + o_b$, and $a_{a_b,b} = a_{A_{b-1},b-1} + a_b$. Hence, the index sets of all sensors and APs in all sub-networks of the whole considered system are $\mathbb{O} = \mathbb{O}_1  ...\cup \mathbb{O}_b .. \cup \mathbb{O}_B =\{1,2,..,o,..,O\}$, and $\mathbb{A} = \mathbb{A}_1 ...\cup \mathbb{A}_b .. \cup \mathbb{A}_B = \{1,2,..,a,..,A\}$, respectively. The total number of sensors is $O= \sum_{b=1}^{B}O_b$ and the total number of APs is $A= \sum_{b=1}^{B}A_b$. Finally, $o$ and $a$ denote the indices of the corresponding sensor and AP within the whole system. % $u_{u_b,b} = u_{U_{b-1},b-1} + u_b$,    $U_b=\#\{\mathbb{U}_b\}$,actuators, $u_b$, actuators, $u_{u_b,b}$, a certain actuator, $\mathbb{U} = \mathbb{U}_1 ...\cup \mathbb{U}_b .. \cup \mathbb{U}_B = \{1,2,..,u,..,U\}$,  $u$, actuators,

  %We always assume that there is an AP collocated with the CU at the center of every subnetwrk. 
Consider that $\boldsymbol{h}_{o,a}\in \mathbb{C}^{M_{a_b}\times 1}$ is the vector containing the channel coefficients from the $o^{th}$ sensor to the $a^{th}$ AP. Particularly, $\boldsymbol{h}_{o,a}$ can be represented as follows. 
\begin{equation}
\label{eq:eq-1}
 \begin{array}{ll}
\boldsymbol{h}_{o,a}& = \sqrt{\beta_{o,a}}\boldsymbol{g}_{o,a},
\end{array}
\end{equation}
where $\beta_{o,a}\in \mathbb{R}$ denotes the large-scale fading factor for the link from the sensor $o$ to the AP $a$. Here, $\beta_{o,a}$ reflects the effects of both the log-normal shadowing along with the path loss of the corresponding link. In addition, the vector $\boldsymbol{g}_{o,a}\in \mathbb{C}^{M_{a_b}\times 1} $ models the small scale Rayleigh fading effect for the link from the sensor $o$ to the AP $a$. Specifically, the vector $\boldsymbol{g}_{o,a} \sim \mathcal{CN}(0,\boldsymbol{I}_{M_{a_b}})$ contains $M_{a_b}$ independent and identically distributed (i.i.d.) Gaussian components, each with zero mean and unit variance. Here, $\boldsymbol{I}_{M_{a_b}}$ represents the $M_{a_b} \times M_{a_b}$ identity matrix.

We consider a time division duplex (TDD) transmission, where the same bandwidth is used for both uplink and downlink transmission.
%Particularly, the coherence interval is divided between these three operations.
In a typical sub-network uplink, there are many sensors sending packets that can originate from simple data recording to high-definition video from cameras.
Conversely, the traffic in the downlink will be commands and control information sent from the CU to the corresponding actuators.
In our work, we focus only on the uplink, assuming that it occupies the channel for a value of time $\tau_{up}$ relative to the channel coherence interval length, while the remaining portion is employed for the downlink transmission. Moreover, in the following Sec.~\ref{Sec:ULRateFormulation} we provide a rate expression of the uplink transmission, considering the distributed MIMO sub-network configuration, together with the constraints introduced by the LBT procedure in the unlicensed spectrum.

% OLD Thus, we focus on the uplink, where we assume that it occupies a value of $\tau_{up}$ relative to the total coherence interval length, while the remaining portion is employed for the downlink transmission. In what follows, we drop the details related to the downlink and elaborate on the uplink transmission of the considered distributed MIMO sub-network, which incorporates communication from sensors to the CU through the APs. %Here, we consider perfect channel knowledge and thus thus will not discuss the details of channel estimation. 

%\section{Compliance with LBT}
Since we are investigating the utilization of the unlicensed band in the context of 6G sub-networks, our design obeys the LBT procedure. %For simplicity,
For this reason, we implement channel sensing at the AP level and consider two transmission schemes. %to ensure compliance with the LBT. 
A baseline scheme, where the transmission in a certain sub-network is suspended/postponed if the average per-antenna uncontrolled interference at any AP belonging to this sub-network is higher than a certain LBT threshold. An adaptive scheme, where we evaluate the impact of an automatic transmission power reduction (APR) approach that decreases the transmission power of every sensor by 1 dB until the interference power at all APs in all the sub-networks is lower than the LBT threshold. Note that the APR is a sort of genie-aided scheme that allows setting the optimal sensor transmit power to make sure that no transmission is suspended/postponed. In that sense, APR represents a bound that allows understanding how beneficial power adaptation mechanisms could be for sub-networks operations in the unlicensed spectrum.

\section{Uplink Rate Formulation} \label{Sec:ULRateFormulation}
In the uplink of a certain sub-network $b$, the sensors send their packets to the APs which convey them to the corresponding CU. These packets can be a  temperature measure, humidity measure, image, video, etc. At a certain AP $a$, the received overall signal, $\boldsymbol{y}_a \in \mathbb{C}^{M_{a_b} \times 1}$, can be expressed by
\begin{equation}
\label{eq:eq-2}
\begin{array}{ll}
\boldsymbol{y}_a& = \displaystyle \sum_{i \in\mathbb{O}} \sqrt{\mu_i}\boldsymbol{h}_{i,a}s_{i}+\boldsymbol{n}_a,
\end{array}
\end{equation} 
where $\mu_i$ is the transmit power used by sensor $i$ and $s_i$ is the data symbol sent by sensor $i$. In addition, $\boldsymbol{n}_a \in \mathbb{C}^{M_{a_b}\times 1}$ is the noise vector at AP $a$, where its $M_{a_b}$ components are i.i.d. additive white Gaussian noise (AWGN). Particularly, $ \boldsymbol{n}_a\sim \mathcal{CN}(0,\sigma_n^2\boldsymbol{I}_{M_{a_b}}) $, where $\sigma_n$ is the standard deviation of the AWGN.
We can rewrite (\ref{eq:eq-2}) so that the desired signal of a certain sensor $o$ is separated from both intra-sub-network interference and inter-sub-network interference at the AP level as follows
%We can recast (\ref{eq:eq-2}) to be as in (\ref{eq:eq-3}) so that the desired signal of certain sensor $o$ can be separable from both intra-sub-network interference and inter-sub-network interference at the AP level. 
\begin{equation}
\label{eq:eq-3}
\begin{array}{ll}
\boldsymbol{y}_a& = \underbrace{\sqrt{\mu_o}\boldsymbol{h}_{o,a}s_{o}}_{\text{Signal of interest at AP}}+ \displaystyle \underbrace{\sum_{i \in\mathbb{O}_b \setminus o} \sqrt{\mu_i}\boldsymbol{h}_{i,a}s_{i}}_{\text{Intra-sub-network interference at AP}}
\\
&+ \underbrace{\displaystyle \sum_{l \in\mathbb{O} \setminus \mathbb{O}_b} \sqrt{\mu_l}\boldsymbol{h}_{l,a}s_{l}}_{\text{Inter-sub-network interference at AP}} +\underbrace{\boldsymbol{n}_a}_{\text{AWGN at AP}},
\end{array}
\end{equation} 
where the received desired packet at AP $a$ from the sensor of interest $o$ is represented by the first term. While the second term indicates the interference at AP $a$ from other sensors in the same sub-network (i.e., the intra-sub-network interference), the third term shows the interference at AP $a$ from sensors belonging to other neighboring sub-networks (i.e., the inter-sub-network interference). Finally,  the last term is the AWGN. 

For uplink signal detection, we assume that all processing is made at the CU.
%Specifically, we deal with each AP as a relay that forward the received packets to the corresponding CU, which will do all the required processing.
Specifically, each AP forwards its received overall signal $\boldsymbol{y}_a$ to the CU belonging to its sub-network that will do all the required processing.
Consequently, the $CU_b$ can construct the corresponding vector of the overall uplink received signal, $\boldsymbol{y}^b\in \mathbb{C}^{M \times 1}$, as follows
\begin{equation}
\label{eq:eq_FC-1}
\begin{array}{ll}
\boldsymbol{y}^b &= \displaystyle \sum_{i \in\mathbb{O}} \sqrt{\mu_i}\boldsymbol{h}_{i}^{b}s_{i}+\boldsymbol{n}^{b},
\end{array}
\end{equation}
where
\begin{equation}
\label{eq:eq_FC-1_2}
\begin{array}{lll}
\boldsymbol{y}^b = \begin{bmatrix}
 \boldsymbol{y}_{a_{1,b}}\\
 :\\
 :\\
\boldsymbol{y}_{a_{A_b,b}}
\end{bmatrix},
&
\boldsymbol{h}_i^b = \begin{bmatrix}
 \boldsymbol{h}_{i,a_{1,b}}\\
 :\\
 :\\
\boldsymbol{h}_{i,a_{A_b,b}}
\end{bmatrix},
&
\boldsymbol{n}^b = \begin{bmatrix}
 \boldsymbol{n}_{a_{1,b}}\\
 :\\
 :\\
\boldsymbol{n}_{a_{A_b,b}}
\end{bmatrix}.
\end{array}
\end{equation}

Let us denote by $\boldsymbol{H}^{b} = [\boldsymbol{h}_{k_b}^b, \boldsymbol{h}_{k_b+1}^b,.., \boldsymbol{h}_{k_b + O_b-1}^b]$ the $\left(M \times O_b\right)$ matrix that represents the $b^{th}$ sub-network overall uplink channel matrix whose columns represent the individual channel vectors from the corresponding sensor in the sub-network $b$ to the APs in the same sub-network.
Here, $k_b=(b-1)O_b+1$ represents the index of the first sensor belonging to the sub-network $b$.

%Furthermore, we can denote the pseudo-inverse matrix of $\boldsymbol{H}^{b}$ by $\boldsymbol{F}^b$, which has the dimension of $(M \times O_b)$ and is given by 
%\begin{equation}
%\label{eq:eq_FC-1_3}
%\begin{array}{ll}
%\boldsymbol{F}^b 
%& = \boldsymbol{H}^{b} \left( \boldsymbol{H}^{b*}\boldsymbol{H}^{b}\right)^{-1},
%\end{array}
%\end{equation} 
%where $(.)^*$ denotes the Hermitian operation, i.e, the conjugate transpose of the operand which can be a vector or a matrix. Thus, the beamforming vector $\boldsymbol{f}_o^b \in \mathbb{C}^{M\times 1}$ for all sensors $o \in \mathbb{O}_{b}$ in the sub-network $b$ can be obtained from the corresponding column in the matrix $\boldsymbol{F}^b$. 

Thus, by denoting with $\boldsymbol{f}_o^b \in \mathbb{C}^{M\times 1}$ the beamforming vector used for sensor $o$,  $CU_b$ multiplies $\boldsymbol{y}^{b}$ by $\boldsymbol{f}^{b*}_{o} $ with $(.)^*$ denoting the Hermitian (conjugate transpose) operation, to get the overall estimate of the desired signal sent by the sensor $o$, $\hat{s}_{o}$, as follows  %& = \boldsymbol{f}^{b*}_{o} \left[ \sqrt{\mu_o}\boldsymbol{h}_{o}^b s_{o}+ \displaystyle \sum_{i \in\mathbb{O}_b \setminus o} \sqrt{\mu_i}\boldsymbol{h}_{i}^b s_{i} \right.\\ & + \left. \displaystyle \sum_{l \in\mathbb{O} \setminus \mathbb{O}_b} \sqrt{\mu_l}\boldsymbol{h}_{l}^b s_{l}+\boldsymbol{n}^b \right], \forall o\in \mathbb{O}_{b} \\
\begin{equation}
\label{eq:eq_FC-2}
\begin{array}{ll}
\hat{s}_{o} & = \boldsymbol{f}^{b*}_{o} \boldsymbol{y}^b, \forall o\in \mathbb{O}_{b}
\\
& =  \underbrace{\sqrt{\mu_o}\boldsymbol{f}^{b*}_{o}\boldsymbol{h}_{o}^b s_{o}}_{\text{Signal of interest at CU}}+\underbrace{ \displaystyle \sum_{i \in\mathbb{O}_b \setminus o} \sqrt{\mu_i} \boldsymbol{f}^{b*}_{o} \boldsymbol{h}_{i}^b s_{i}}_{\text{Intra-sub-network interference at CU}} 
\\
& + 
\underbrace{\displaystyle \sum_{l \in\mathbb{O} \setminus \mathbb{O}_b} \sqrt{\mu_l} \boldsymbol{f}^{b*}_{o}\boldsymbol{h}_{l}^b s_{l}}_{\text{Inter-sub-network interference at CU}} + \underbrace{ \boldsymbol{f}^{b*}_{o}\boldsymbol{n}^b }_{\text{AWGN at CU}} , \forall o\in \mathbb{O}_{b}.
\end{array}
\end{equation} % \right. \\ &    \left.
where, at the CU level, the first term represents the overall desired signal, the second term the overall intra-sub-network interference, the third term the overall inter-sub-network interference, and the fourth term the overall AWGN. It should be noted that intra-sub-network interference is considered a sort of controlled interference (CI) where the CU of a certain sub-network can remove it for example by employing zero-forcing (ZF) beamforming towards the sensors within the same sub-network:  more specifically with ZF, beamforming vector $\boldsymbol{f}_o^b$ is obtained from the corresponding column of matrix $\boldsymbol{F}^b = \boldsymbol{H}^{b} \left( \boldsymbol{H}^{b*}\boldsymbol{H}^{b}\right)^{-1}$.
%Specifically, the CU of every sub-network can utilize the estimated channels from the antennas of all APs in the sub-network to all sensors in the sub-network to design the ZF beamforming vectors.
On the other hand, inter-sub-network interference is an uncontrolled interference (UI) as the CU of a certain sub-network can not control the interference from the sensors belonging to neighboring sub-networks.

%\section{URLLC Rate Expression}
From equation (\ref{eq:eq_FC-2}), we observe that the overall received desired signal power for sensor $o$, $P^{s}_o$, is expressed as 
\begin{equation}
\label{eq:eq-PF_1_FCU}
\begin{array}{ll}
P^{s}_{o} 
& = \mu_o \left | \displaystyle  \boldsymbol{f}^{b*}_{o} \boldsymbol{h}_{o}^b\right |^2.
\end{array}
\end{equation} 
Also, the overall CI power for sensor $o$, $P^{CI}_o$, is given by
\begin{equation}
\label{eq:eq-PF_2_FCU}
\begin{array}{ll}
P^{CI}_{o} 
& = \displaystyle \sum_{i \in\mathbb{O}_b \setminus o} \mu_i \left |  \boldsymbol{f}^{b*}_{o}\boldsymbol{h}_{i}^b\right |^2,
\end{array}
\end{equation} 
and the overall UI power for sensor $o$, $P^{UI}_o$, is given by 
\begin{equation}
\label{eq:eq-PF_2a_FCU}
\begin{array}{ll}
P^{UI}_{o} 
& = \displaystyle \sum_{l \in\mathbb{O} \setminus \mathbb{O}_b} \mu_l \left | \boldsymbol{f}^{b*}_{o}\boldsymbol{h}_{l}^b\right |^2.
\end{array}
\end{equation} 

Thus, the signal to interference plus noise ratio (SINR) for sensor $o$ can be expressed by %equation (\ref{eq:eq-PF_3_FCU})
\begin{equation}
\label{eq:eq-PF_3_FCU} 
\begin{array}{ll}
\Gamma_o& = \displaystyle \frac{\mu_o \left |  \boldsymbol{f}^{b*}_{o} \boldsymbol{h}_{o}^b\right |^2}{\displaystyle \sum_{i \in\mathbb{O}_b \setminus o} \mu_i \left |  \boldsymbol{f}^{b*}_{o}\boldsymbol{h}_{i}^b\right |^2+ \displaystyle \sum_{l \in\mathbb{O} \setminus \mathbb{O}_b} \mu_l \left | \boldsymbol{f}^{b*}_{o}\boldsymbol{h}_{l}^b\right |^2 + \sigma_n^2  \left\| \boldsymbol{f}^{b*}_{o}\right\|^2}.
\end{array}
\end{equation}
By employing the normal approximation of the finite blocklength capacity \cite{durisi16}, the data rate for the sensor $o$ is given by %\cite{polyanskiy2010channel, mary2016finite, nasir2020resource}
\begin{equation}
\label{eq:eq-PF_4}
\begin{array}{ll}
R_o& = \tau_{up}W \left ( \vphantom{\displaystyle \frac{Q^{-1}(\zeta)}{\sqrt{DW}}}  {\log}_2 (1+ \Gamma_o) - \left( 1-\frac{1}{(1+ \Gamma_o)^2}\right)^{1/2}   \right.
\\
& . \left. 
 \displaystyle \frac{Q^{-1}(\zeta)}{\sqrt{DW}}\log_2 \left(e \right) \right ) \quad{\mbox{bit/sec}}, 
\end{array}
\end{equation} %\right.  \\ &%\right.  \\ &  \left. \times
%\\& = B \left ( log_2 (1+ \gamma_u^{dl}[n]) - \frac{\sqrt{\gamma_u^{dl}[n](\gamma_u^{dl}[n]+2)}}{1+\gamma_u^{dl}[n]}\frac{Q^{-1}(\epsilon)}{\sqrt{tB}}log_2 (e)\right ) bits/sec, %$W$ denotes transmission bandwidth,
where $\zeta$ defines the probability of packet error, $e$ represents the natural exponent, $D$ indicates packet transmission duration and $Q^{-1}(.)$ is the inverse of the complementary cumulative distribution function of the standard Gaussian random variable. Note that equation (\ref{eq:eq-PF_4}) is the Shannon capacity corrected by the channel dispersion term. Note also that (\ref{eq:eq-PF_4}) is the rate for the sensors belonging to the sub-networks that successfully performed the LBT: in our model, when a sub-network does not pass the LBT, we simply assume no rate for its sensors.

\section{Numerical Results}
We consider an indoor factory propagation environment with dense clutter and low AP heights (InF-DL), where the effective clutter height is 10 m and the clutter size is 2 m \cite{3rd20223gpp}. We assume a square shaped factory hall with a side length of 100 m and a ceiling height of 15 m. We assume that sub-networks are uniformly distributed in the considered factory environment such that the distance between the centers of any two neighboring sub-networks is at least 10 m. In addition, every sub-network is assumed to have a radius of 5 m, wherein the control unit is located at its center at the height of 5 m. We assume that there is an AP at the CU location, whereas the remaining $A_b-1$ APs are uniformly distributed within the corresponding sub-network at a height of 5 m. Also, we assume that a number of $O_b = 5$ sensors are uniformly distributed within each sub-network at a height of 1.5 m. 

Furthermore, we consider a channel bandwidth $W=100$ MHz in the 6 GHz unlicensed band, a packet transmission duration $D=50 \mu $s, and a target packet error rate $\zeta = 10^{-6}$. We assume that the proportional uplink transmission period relative to the coherence interval is $\tau_{up}=0.4$. The receiver noise figure is  9 dB, and the noise power spectral density is -174 dBm/Hz. The LBT threshold is assumed to be -72 dBm. In our experiments, we assume each sub-network is served with a total number of $M=20$ antennas equally distributed among the deployed APs, and we compare the performance of three different cases of antenna distribution. Namely, we consider $A_b \in \{20,5,1\}$ and in every case, the number of antennas at every AP is given by $M_{a_b} = \frac{M}{A_b} \in\{1, 4, 20\}$. 

\begin{figure}[!t]
\begin{center}
\includegraphics[width = 1\columnwidth]{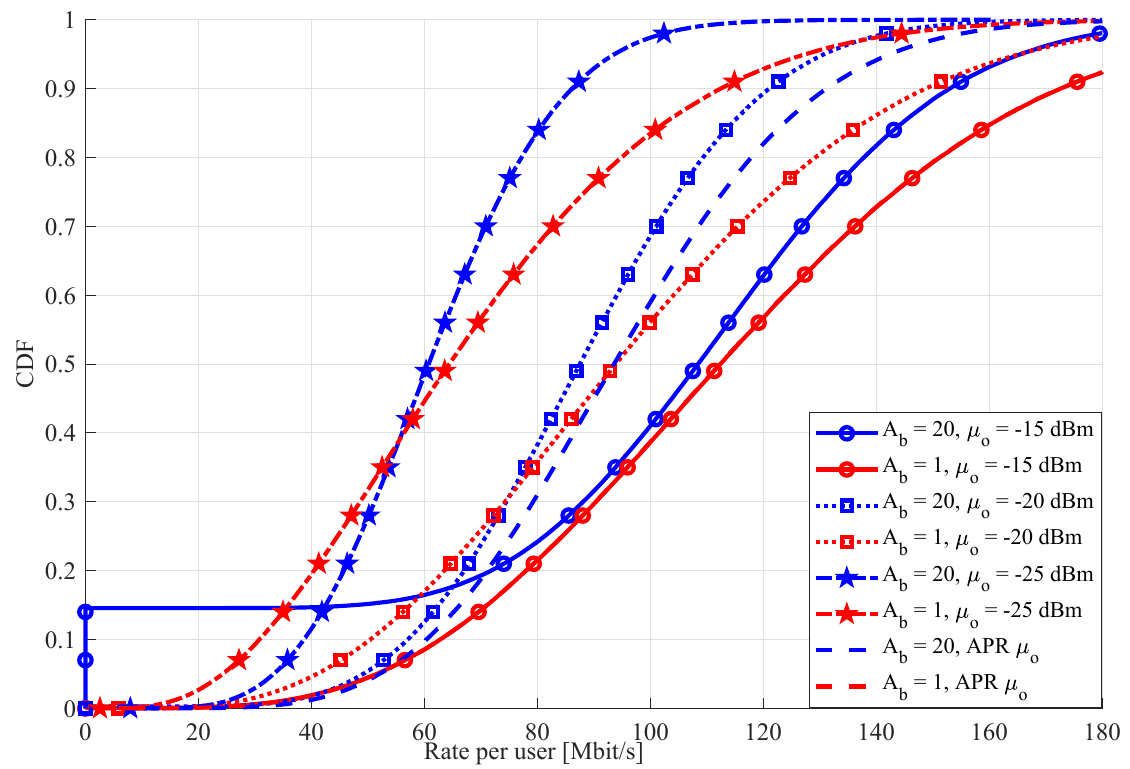}
\caption{CDF of the uplink rate comparing APR and the baseline with $\mu_o \in \{-15 {\rm dBm}, -20 {\rm dBm}, -25 {\rm dBm}\}$ when $B = 30$.} %\vspace{-.2in}
\label{fig:rate_15_20_25_optimizedmu_5Ob_30B}
\end{center}
\end{figure}
\begin{figure}[!t]
\begin{center}
\includegraphics[width = 1\columnwidth]{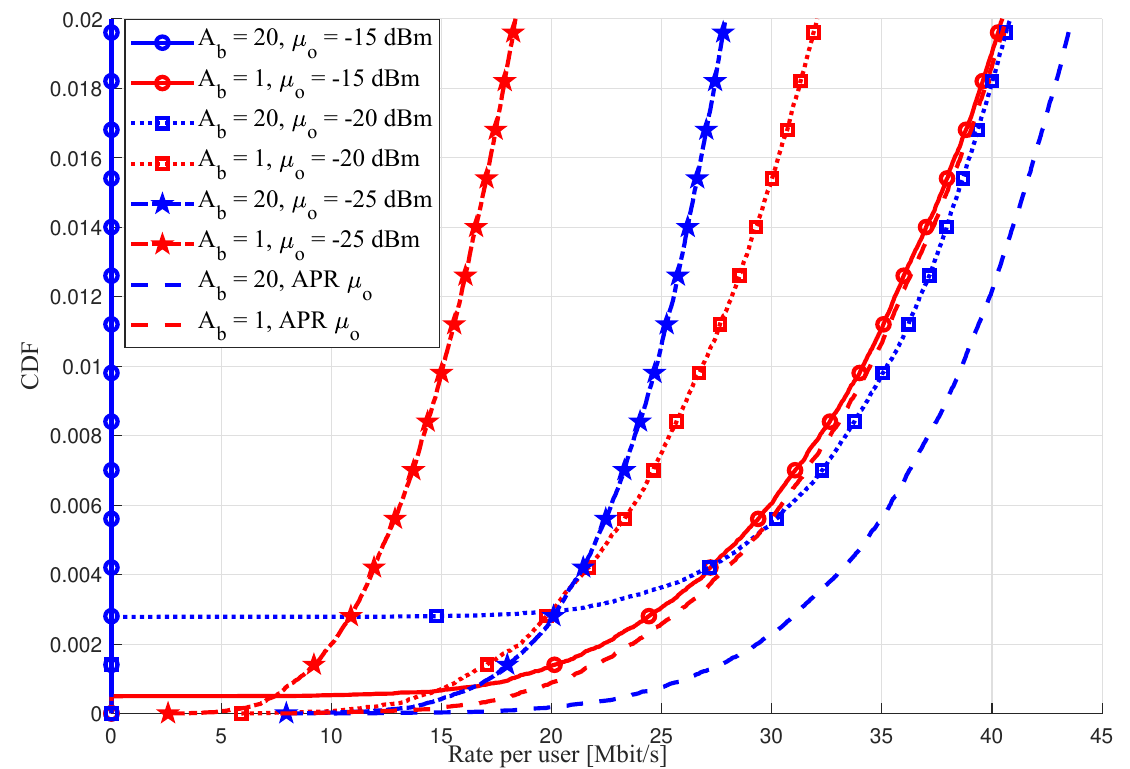}
\caption{Zoom on the lower part of the CDF curves in Fig. \ref{fig:rate_15_20_25_optimizedmu_5Ob_30B}.} %\vspace{-.2in}
\label{fig:zoom_rate_15_20_25_optimizedmu_5Ob_30B}
\end{center}
\end{figure}
%In this part, we show the motivation for employing the APR scheme to avoid the need for deferring the transmission of any sub-network. Specifically,
Fig. \ref{fig:rate_15_20_25_optimizedmu_5Ob_30B} shows the cumulative distribution function (CDF) of the sensor uplink rate for the cases of baseline backoff scheme with sensor fixed transmit power $\mu_o \in \{-15{\rm dBm}, -20 {\rm dBm}, -25 {\rm dBm}\}$ against the APR scheme for a system with $B=30$ sub-networks for both cases of fully distributed MIMO ($A_b = 20$) and fully centralized MIMO ($A_b = 1$).
In addition to that, Fig. \ref{fig:zoom_rate_15_20_25_optimizedmu_5Ob_30B} makes a zoom in the lower part around the 0.01 CDF values of Fig. \ref{fig:rate_15_20_25_optimizedmu_5Ob_30B} curves.
% We perform simulations for 2000 iterations for every case.
In the case of fully distributed MIMO system, i.e., when $A_b=20$, the percentage of sensors with no transmission falls from about 15\% with $\mu_o = -15 {\rm dBm}$ to about 0.3\% and 0\% with $\mu_o = -20{\rm dBm}$ and $\mu_o = -25 {\rm dBm}$, respectively. Moreover, the improvements between the fully distributed approach, i.e., $A_b=20$, and the centralized one, i.e., $A_b=1$, are evident, except for the case when $\mu_o = -15 {\rm dBm}$ where some sub-networks suffer from UI from neighboring sub-networks and defer their transmissions.
%For both cases when $\mu_o = -20 dBm$ and  $\mu_o = -25 dBm$, performance of the fully distributed MIMO system, i.e, when $A_b=20$, the CDF performance is better compared to the corresponding counterparts at $A_b=1$. Specifically, at the small values of the CDF, the rates of $A_b=20$ are much higher than their counterparts of $A_b=1$ when  $\mu_o = -20 dBm$ and  $\mu_o = -25 dBm$. On the other hand, when $\mu_o = -15 dBm$, there will be some sub-networks that suffer from UI from neighboring sub-networks, and thus, these sub-networks defer their transmission leading to an impact on the uplink sensor performance. 
It is clear that the impact of deferring the transmission increases with the sensor transmission power, especially when the antennas are distributed. On the other hand, we highlight that the APR scheme is able to deal with that issue as the performance of the APR is better when compared to the baseline backoff scheme at the different static sensor transmit powers, regardless of the number of APs in the sub-network. Furthermore,
%it is noteworthy that part of the CDF plots is the most important where it reflects the performance of the unfortunate sensors that has the worst channel conditions.
Fig. \ref{fig:zoom_rate_15_20_25_optimizedmu_5Ob_30B} depicts that, when the APR is utilized, the uplink rate of the sensors with worst channel conditions, i.e., in the lower part of the CDF, has the best performance in the case of fully distributed MIMO case. Compared to centralized MIMO, the employment of the APR along with the fully distributed MIMO guarantees a better rate for sensors with worst channel conditions and consequently improves the worst case delay.

\begin{figure}[!t]
\begin{center}
\includegraphics[width = 1\columnwidth]{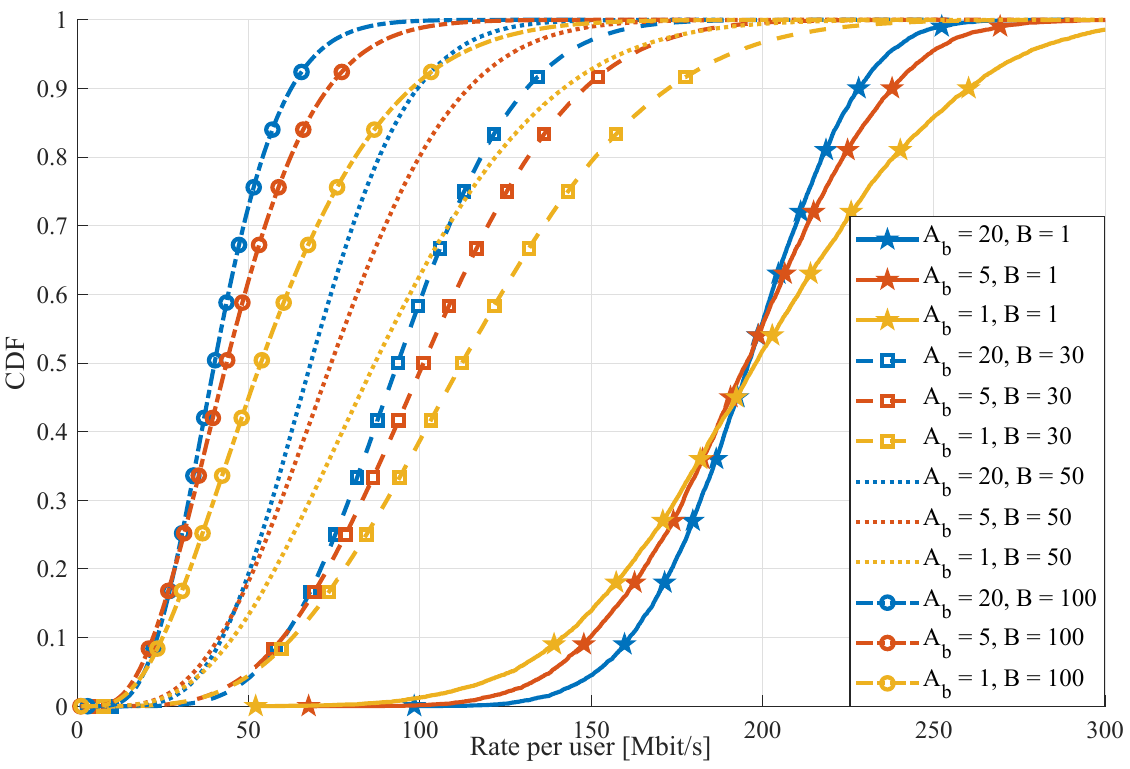}
\caption{CDF of the uplink rate for APR with $B \in \{1,30,50,100\}$.} %\vspace{-.2in}
\label{fig:rate_Optimizedmu_5Ob_1_30_50_100B}
\end{center}
\end{figure} 

\begin{table} [!t]
\centering
\caption{Comparison of the uplink rate of both 0.01 and 0.001 CDF at different antenna distribution levels with different numbers of sub-networks.}
\resizebox{\columnwidth}{!} {
\begin{tblr}{
  cell{1}{1} = {r=2}{},
  cell{1}{2} = {c=3}{},
  cell{1}{5} = {c=3}{},
  vlines,
  hline{1,3-7} = {-}{},
  hline{2} = {2-7}{},
}
        & Rate [Mbit/s] of the 0.01 CDF &         &         & Rate [Mbit/s] of the 0.001 CDF &         &         \\
        & $A_b=20$                      & $A_b=5$ & $A_b=1$ & $A_b=20$                       & $A_b=5$ & $A_b=1$ \\
$B=1$   & 132.58                        & 116.87  & 96.38   & 114.63                         & 96.96   & 68.41   \\
$B=30$  & 38.64                         & 34.54   & 34.46   & 26.12                          & 21.43   & 20.51   \\
$B=50$  & 26.43                         & 23.08   & 23.9    & 17.94                          & 14.16   & 13.82   \\
$B=100$ & 13.95                         & 11.59   & 12.22   & 9.26                           & 6.91    & 6.96    
\end{tblr}
}
\label{tab:RateComparison}
\end{table}
In  Fig. \ref{fig:rate_Optimizedmu_5Ob_1_30_50_100B}, we study the impact of the number of sub-networks on the sensor uplink rate when the APR scheme is employed by considering $B \in \{1,30,50,100\}$.
%Now, we concentrate more on the performance of the APR. Specifically, we study the impact of the number of sub-networks on the individual sensor uplink performance when the APR scheme is employed.
%Specifically, Fig. \ref{fig:rate_Optimizedmu_5Ob_1_30_50_100B} indicates the CDF plots for the uplink rates performance of the APR at different values of $B \in \{1,30,50,100\}$ when $O_b = 5$.
Note that the case $B=1$ represents the interference-free scenario which corresponds to the upper bound, and the case $B=100$ can be considered as the worst-case scenario where all neighboring sub-networks are contiguous to each other. As expected, increasing the number of sub-networks increases the UI levels for every sensor in the system, which consequently reduces the sensor uplink rate performance regardless of the antenna distribution level. On the other hand, for a certain number of sub-networks and when looking at small CDF values below 0.01, the APR uplink rate performance of the fully distributed MIMO, i.e., with $A_b=20$, presents an advantage when compared to those of  $A_b=5$ and $A_b=1$. More in details, Table \ref{tab:RateComparison} presents the rate comparison with APR at the CDF values of 0.01 and 0.001 between the considered three levels of antenna distribution and the considered four different numbers of sub-networks. It is clear that the fully distributed MIMO, i.e., when $A_b = 20$, always provides the highest 0.01 CDF rate and the highest 0.001 CDF rate compared to both cases when $A_b = 5$ and $A_b = 1$ regardless of the number of sub-networks in the considered system.
%At the median of the CDF, the centralized approach, i.e., $A_b=1$, returns the best uplink rates, regardless of the sub-network density. 

\begin{figure}[!t]
\begin{center}
\includegraphics[width = 1\columnwidth]{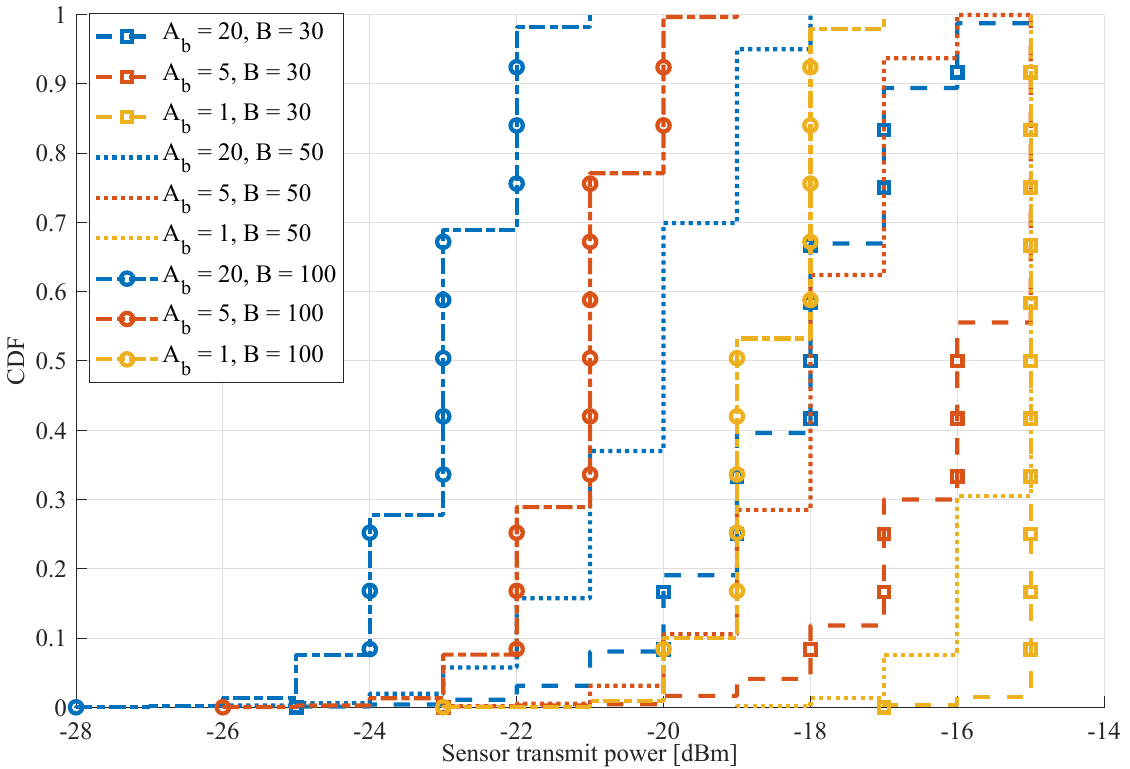}
\caption{CDF of the sensor transmit power for APR with $B \in \{30,50,100\}$.} %\vspace{-.2in}
\label{fig:Power_Optimizedmu_5Ob_30_50_100B}
\end{center}
\end{figure}
Finally, Fig. \ref{fig:Power_Optimizedmu_5Ob_30_50_100B} shows the CDF of the sensor transmit power for different values of $B \in \{30,50,100\}$ when the APR is employed. We notice that increasing the number of sub-networks in the service area needs more reduction in the sensor transmit power to avoid deferring the transmission within any sub-network. This is justified because increasing the number of sub-networks means that the UI is increased and thus we need to reduce the sensor transmit power further to guarantee that there is no sub-network having UI higher than the LBT threshold.  This ensures that none of the sub-networks will need to defer its transmission. Moreover, the reduction of the transmitting power has clear advantages in terms of energy efficiency and battery life duration of the sensors.

\section{Conclusions}
In this paper, we have investigated the potential of employing the 6 GHz unlicensed
spectrum to support 6G sub-networks use cases.
Moreover, we have assessed the benefits of using distributed MIMO to improve performance.
%In this paper, we have investigated the potential of employing both the distributed MIMO along with the 6 GHz unlicensed spectrum to support URLLC services for the 6G sub-networks.
%We employed ZF beamforming to remove the CI from other sensors within the same sub-network.
As the UI from other sensors in the neighboring sub-networks limits the possibility of accessing the channel when using LBT, we have presented the APR scheme that reduces the transmit power to increase the likelihood of accessing the spectrum.
%Furthermore, we presented the APR scheme to cope with the UI from other sensors in the neighboring sub-networks, that limits the possibility to access the channel because of the LBT protocol.
Our numerical results verify the effectiveness of utilizing APR along with distributed MIMO over the unlicensed spectrum to ensure the critical QoS requirements are met for reliable operations of the future sub-networks. \balance
%Finally, this paper opens the door for further investigations into the role of unlicensed spectrum in future sub-networks.

\bibliographystyle{IEEEtran}
\bibliography{bibfile}
\end{document}